\documentclass[12pt,preprint,apj]{emulateapj}

\shorttitle{Globular Clusters in the M81 Group}
\shortauthors{Jang et al.}
\begin{document}

\title{Discovery of the Most Isolated Globular Cluster in the Local Universe }

\author{In Sung  Jang, Sungsoon Lim, Hong Soo Park, and Myung Gyoon Lee}
\affil{Astronomy Program, Department of Physics and Astronomy, Seoul National University, Gwanak-gu, Seoul 151-742, Republic of Korea}
\email{isjang@astro.snu.ac.kr, slim@astro.snu.ac.kr, hspark@astro.snu.ac.kr, and mglee@astro.snu.ac.kr}


\begin{abstract}

We report the discovery of  two new globular clusters in the remote halos of M81 and M82 in the M81 Group based on the Hubble Space Telescope archive images. 
They are brighter  than typical globular clusters 
($M_{V}=-9.34$ mag for GC-1 and $M_{V}=-10.51$ mag for GC-2),
and much larger than known globular clusters with similar luminosity in the Milky Way Galaxy and M81. 
Radial surface brightness profiles for GC-1 and GC-2  do not show any feature for tidal truncation in the outer part.
They are located much farther from either of M81 and M82 in the sky, compared with previously known star clusters in these galaxies. 
Color-magnitude diagrams of resolved stars in each cluster show a well-defined red giant branch (RGB), indicating that they are metal-poor and old. 
We derive a low metallicity with [Fe/H] $\approx -2.3$
and an old age $\sim 14$ Gyr for GC-2 from the analysis of the absorption lines in its spectrum in the Sloan Digital Sky Survey in comparison with the simple stellar population models.  
The $I$-band magnitude of the tip of the RGB for GC-2 is 0.26 mag fainter than that for the halo stars in the same field,  
showing that GC-2 is $\sim$400 kpc behind the M81 halo along our line of sight.
The deprojected distance to GC-2 from M81 is much larger than any other known globular clusters in the local universe.
This shows that GC-2 is the most isolated globular cluster in the local universe.  

\end{abstract}



\section{Introduction}

According to the current paradigm of large scale structure formation, individual galaxies, galaxy clusters, and groups are formed via hierarchical merging of galaxies.
Globular clusters are a powerful tool to test this hypothesis.
Globular clusters are often found in and around galaxies.
With the advent of wide field surveys, globular clusters are sometimes found in the remote halo of galaxies in the Local Group. A small number of globular clusters are found beyond 30 kpc from the center of the Milky Way Galaxy (MWG) and M31 \citep{har96,hux08, gal07},  while most of the globular clusters are much closer to the galaxy center. A few clusters are found also in the remote halo of less massive galaxies such as NGC 6822 \citep{hwa11} and M33 \citep{sto08,coc11}.
Recently intracluster globular clusters are also found in nearby galaxy clusters: Virgo \citep{wil07, lee10}, Coma \citep{pen11}, and Abell 1835 \citep{wes11}.
Globular clusters in galaxies provide fossil records for early formation of spheroidal components in the collapsing phase, while globular clusters far from galaxies reveal clues for later growth of galaxies via accretion.

The M81 Group, one of the nearest galaxy groups, is an excellent laboratory for studying the property of dwarf galaxies and star clusters as well as starburst galaxies and intergalactic medium \citep{chi09}. The main galaxy located in the center of the group is M81, surrounded by 29 member galaxies \citep{chi09,mak11}. 
According to \citet{mak11},  the systemic velocity with respect to the Local Group and the velocity dispersion of the M81 Group are $v_{LG} = 193 $ km s$^{-1}$  and $\sigma_v=138$ km s$^{-1}$, respectively, and the total mass derived from
the velocity of the members is $M= 3.89  \times 10^{12} M_\odot $. 
The M81 Group includes also a famous starburst galaxy M82. 
\citet{kar06} estimate from kinematics and distances of the member galaxies that
M81 is twice as massive as M82 and the mass of the M81 Group is 77$\%$ of the mass of the Local Group. 

The distance to M81 is known to be $3.63 \pm 0.14$ Mpc ($(m-M)_0= 27.80 \pm 0.08$) derived using the tip of the red giant branch (TRGB) method from deep HST images as well as Cepheids by \citet{dur10}. 
The distance to M81 is estimated to be similar to that to M82 based on the same TRGB method,  $3.55\pm 0.11$ Mpc ($(m-M)_0= 27.75\pm 0.07$) \citep{lee12}, 
showing that these two galaxies are at a similar distance from us.

Previous studies found numerous star clusters in M81 and M82, which are mostly located in the main body of each galaxy \citep{may08, nan10, nan10b, san10, san11, nan11}. 
We have been searching for globular clusters in  a remote halo region of each galaxy, much farther
from previous surveys.
In this paper we present  a discovery of two globular clusters in the remote region of M81 and M82.
This paper is composed as follows.
In Section 2, we describe data used and the globular  cluster search method.
Section 3 presents the discovery of two new globular clusters,
 the color-magnitude diagrams of resolved stars in each cluster, and
distance measurements. We also present the measurements of metallicity, age, and [$\alpha$/Fe] using the spectrum of one globular cluster.
Finally we derive the surface brightness profiles of the new globular clusters.
Primary results are discussed and summarized in the final section.

\section{Data and Cluster Search} 

We used the images taken with the   Hubble Space Telescope (HST)/Advanced Camera for Surveys (ACS) and Wide Field Camera 3 (WFC3) in the archive to search for globular clusters in the remote halo of M81 and M82. 
There are $\sim$50 sets of multi-band images for the remote regions in the M81 group available. 
The images in which we found new globular clusters are  ACS/WFC $F606W$ and $F814W$ images (obtained with exposure times 850 s and 690 s, respectively),
and  WFC3/UVIS $F606W$ and $F814W$ images (obtained with exposure times 735 s and 1225 s, respectively).
Both fields come from the HST program 11613 (P.I. : de Jong).

We expected that globular clusters far from galaxies are more extended than those close to galaxy centers, so that
some stars in the outer region of these globular clusters at the distance of the M81 Group can be resolved in the HST images. 
We searched all these fields visually with a primary aim to find globular clusters.
To identify an object as a globular cluster, we used the following visual criteria : (1) round appearance, (2) lack of a smooth extended halo typical of background elliptical galaxies, and (3) being surrounded by an excess of resolved stars.

\begin{figure}
\centering
\includegraphics[scale=0.9]{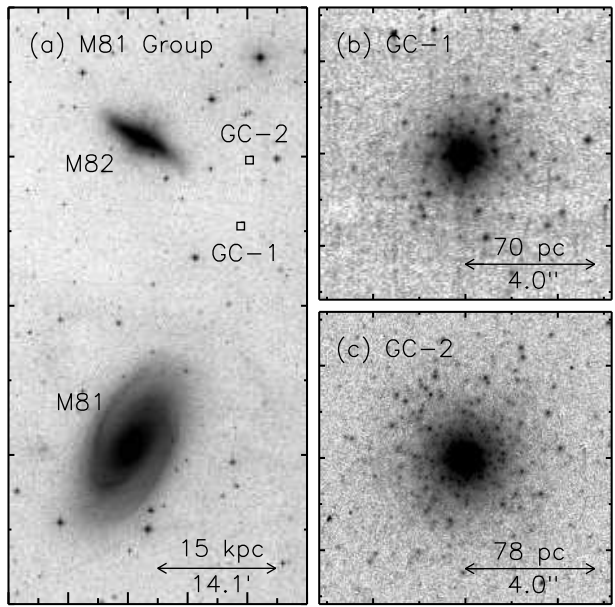} 
\caption{(a) A gray scale map of the digitized sky survey for the central region of the M81 Group, 
showing the positions of GC-1 and GC-2 discovered in this study. North is up and east to the left. 
(b) and (c) Gray scale maps of the $F814W$ images for GC-1 and GC-2.} 
\label{figmap}
\end{figure}

\section{Results}
 
\subsection{Discovery of New Globular Clusters}

Through visual search 
we discovered two new globular clusters, JM81GC-1 and JM81GC-2 (called GC-1 and GC-2 hereafter), in ACS/WFC and WFC3/UVIS fields, respectively.
Figure \ref{figmap}(a) shows the location of these globular clusters. 
GC-1 is 15$\arcmin$.90 west of M82 and 29$\arcmin$.58 north of M81 
(corresponding projected distances are 17.07 kpc and 31.77 kpc, respectively).
GC-2 is 13$\arcmin$.34 west of M82 and 37$\arcmin$.25 north of M81 
(corresponding projected distances are 14.32 kpc and 40.00 kpc, respectively).
Their positions are 
RA(2000)=09$^h$ 53$^m$ 26.22$^s$, Dec(2000)=69$^\circ$ 31$\arcmin$ 17$\arcsec$.5  for GC-1, and 
RA(2000)=09$^h$ 53$^m$ 20.17$^s$, Dec(2000)=69$^\circ$ 39$\arcmin$ 16$\arcsec$.4   for GC-2.

We checked the images of these clusters in the Sloan Digital Sky Survey (SDSS) \citep{yor00}. 
GC-1 and GC-2 appear as extended source in the SDSS images and were classified as galaxies (SDSS ID : J095326+693117 for GC-1 and J095320+693916 for GC-2). 
However, $F814W$ images in Figure 1(b) and (c) show some resolved stars in the outer region of these clusters, proving that they are genuine star clusters. 
We checked the existence of any nearby dwarf galaxies around these two clusters using the data in \citet{chi09} as well as the HST images, but found none.

\begin{figure*}
\centering
\includegraphics[scale=0.80]{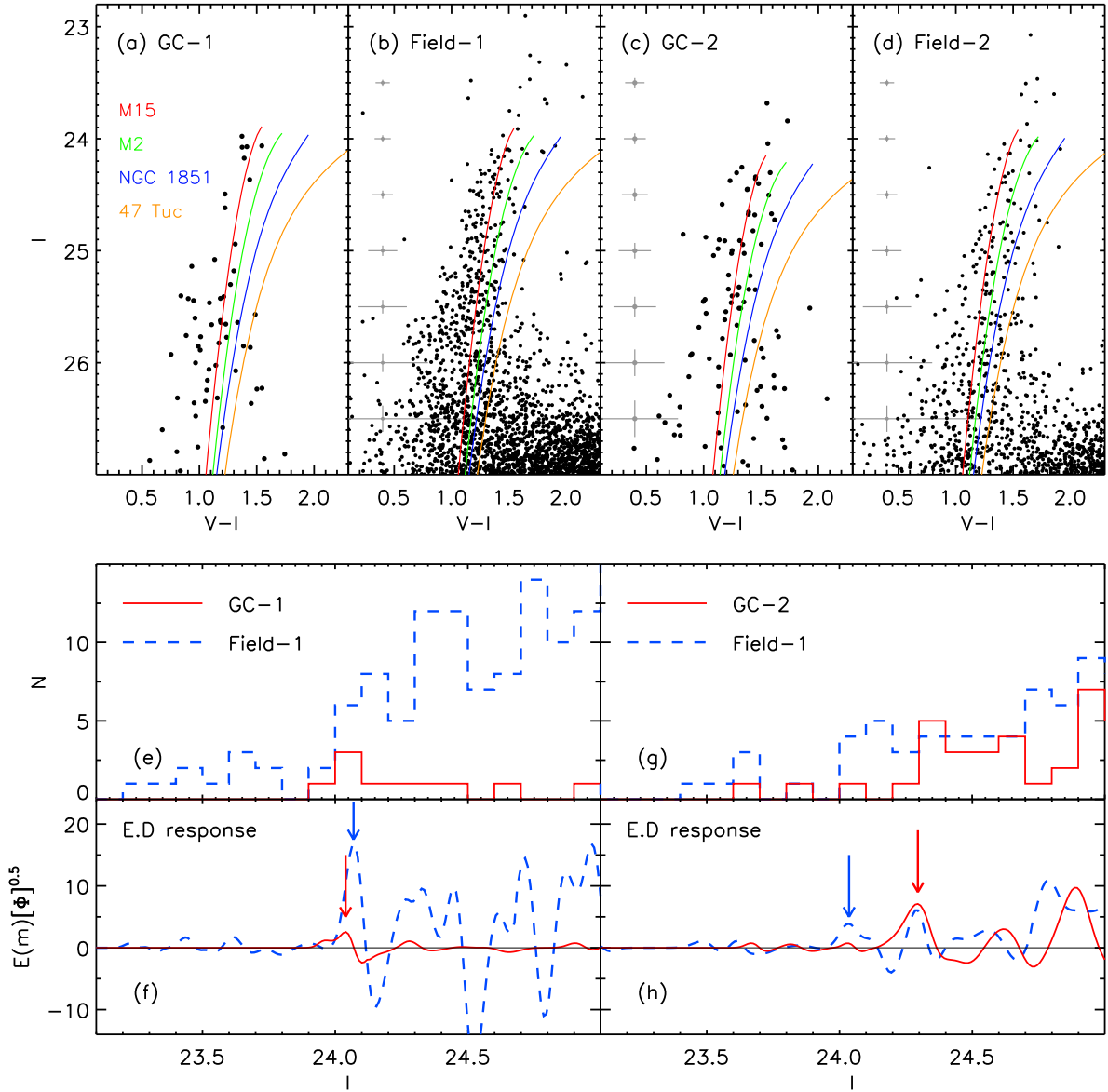}
\caption{(a)$-$(d) $I$--$(V-I)$ color-magnitude diagrams of the stars in the region at $0\arcsec.8<r<6\arcsec.0$  for GC-1 and GC-2 and in their corresponding fields at $15\arcsec <r\lesssim100\arcsec$. 
The curved lines represent the loci of the RGB in the Milky Way globular clusters  with a range of metallicity
 ([Fe/H] = --2.17,  --1.58,  --1.29, and --0.71, respectively),
 shifted according to the foreground reddening and derived distance. Errorbars represent the mean errors.
(e)--(h) $ I$-band luminosity functions of red giants with $0.7 < (V-I) < 1.9$ (e and g), and weighted edge-detection responses (f and h).
Arrows represent the positions of the TRGB.
}
\label{figcmd}
\end{figure*}

\subsection{Color-Magnitude Diagrams of Resolved Stars}
 
We derived instrumental magnitudes of point sources in the images using the IRAF/ DAOPHOT package
 that is designed for point spread function (PSF) fitting photometry \citep{ste94}.
We used 2-$\sigma$ as the detection threshold, and derived the PSFs using isolated bright stars in the images. We applied aperture correction derived
from isolated bright stars.  
We calibrated the instrumental magnitudes to the Vega magnitude system using photometric zeropoints for ACS/WFC (http://www.stsci.edu/hst/acs/analysis/zeropoints/$\#$
tablestart) and WFC3/UVIS (http://www.stsci.edu/
hst/wfc3/phot$\_$zp$\_$lbn). Then we converted this system to Johnson-Cousins $VI$ system using \citet{sir05}.

Figures \ref{figcmd}(a)$-$(d) display the 
color-magnitude diagrams for GC-1 and GC-2 as well as
corresponding fields
(called Field 1 and Field 2, respectively). Because the crowding is severe in the central region
of the globular clusters, we plotted the stars at $0\arcsec.8 <r< 6\arcsec.0$  from the cluster center. 
The aperture with radius of 6$\arcsec$ covers about 96 \% of the total luminosity of each star cluster.
In the case of fields, we plotted the stars at  $15\arcsec <r\lesssim100\arcsec$.
Both clusters show a relatively well-defined red giant branch
(RGB), indicating that they may be old globular clusters.

\subsection{Distance Estimation}
 
We derived the distance to these clusters as well as to the corresponding halos using the TRGB method \citep{lee93,sak96,men02,mcc04,mou10,con11}.
In Figure \ref{figcmd}(e) the $I$-band luminosity function of the red giants in GC-1 and Field-1 shows a sudden jump at $I \approx 24.0$, which corresponds to the TRGB. 
However, the TRGB magnitude for GC-2 is about 0.3 mag fainter than that for Field-2 in Figure \ref{figcmd}(g), showing that
GC-2 may be behind the halo stars. 

Using the edge-detecting algorithm,  we determined the TRGB magnitude
more quantitatively.
 We calculated an edge-detection response function 
$E(m)$ ($= \Phi (m + \sigma_m ) -   \Phi (m - \sigma_m )  $ where $\Phi (m)$ is the luminosity function
of magnitude $m$ and $\sigma_m$ is the mean photometric error within a bin of $\pm0.05$ mag about magnitude $m$), and we   weighted it 
according to the Poisson noise of the luminosity function $E(m)\sqrt{\Phi (m) }$ \citep{men02}, as shown in Figures \ref{figcmd}(f) and (h).

Thus derived TRGB magnitudes are 
$I_{\rm TRGB}=24.039\pm0.021$ for GC-1,        $24.069\pm0.011$ for Field-1, 
                 $24.295\pm0.036$ for GC-2,  and $24.036\pm0.022$ for Field-2. 
The errors for the TRGB magnitudes were 
determined using bootstrap resampling method with one million simulations. In each simulation, we resampled randomly
the RGB sample with replacement to make a new sample of the same size. We estimated the TRGB magnitude for
each simulation using the same procedure and derived the standard deviation of the estimated TRGB magnitudes.

The mean color of the TRGB is derived from the colors of the bright red giants close to the TRGB and is corrected for the foreground reddening ($E(B-V)$ = 0.09 for GC-1 and $E(B-V)$ = 0.10 for GC-2 \citep{sch98}):
 $(V-I)_{0,\rm TRGB}=1.32 \pm0.02 $ for GC-1,   
 $1.41 \pm0.03 $ for Field-1, 
 $1.31 \pm0.04 $ for GC-2, 
 and $1.49 \pm0.04 $ for Field-2. 
 We derive the intrinsic $I$-band magnitude of the TRGB using the relation between the bolometric magnitude $M_{\rm bol}$ and the bolometric correction
$BC$ : $M_{\rm I_0}$ = $M_{\rm bol} - BC_{\rm I_0}$.
We calculate the bolometric magnitude using $M_{\rm bol}$ = $-0.19 {\rm [Fe/H]} - 3.81$
 and the bolometric correction using   $BC_{\rm I_0} = 0.881 - 0.243(V -I)_{\rm 0,TRGB}$ \citep{dac90}.
[Fe/H] is derived from the mean color of the RGB stars. 

The mean color of the RGB stars 0.5 mag fainter than the TRGB is derived from the colors of the bright red giants close to this magnitude:
 $(V-I)_{0, -3.5}=1.21 \pm0.02 $ for GC-1,  
 $1.22 \pm0.03 $ for Field-1, 
 $1.21 \pm0.03 $ for GC-2, and 
 $1.31 \pm0.02 $ for Field-2. 
 From these we derive [Fe/H]:
 [Fe/H] = $-2.23\pm0.11$ for GC-1, 
 $-2.18\pm0.13$ for Field-1, 
 $-2.23\pm0.12$ for GC-2, and
$-1.84\pm0.10$ for Field-2.
Using these we obtain 
 $M_{\rm I,TRGB}=-3.95 $ for GC-1, $-3.93$
 
 \begin{figure*}
\centering
\includegraphics[scale=0.8]{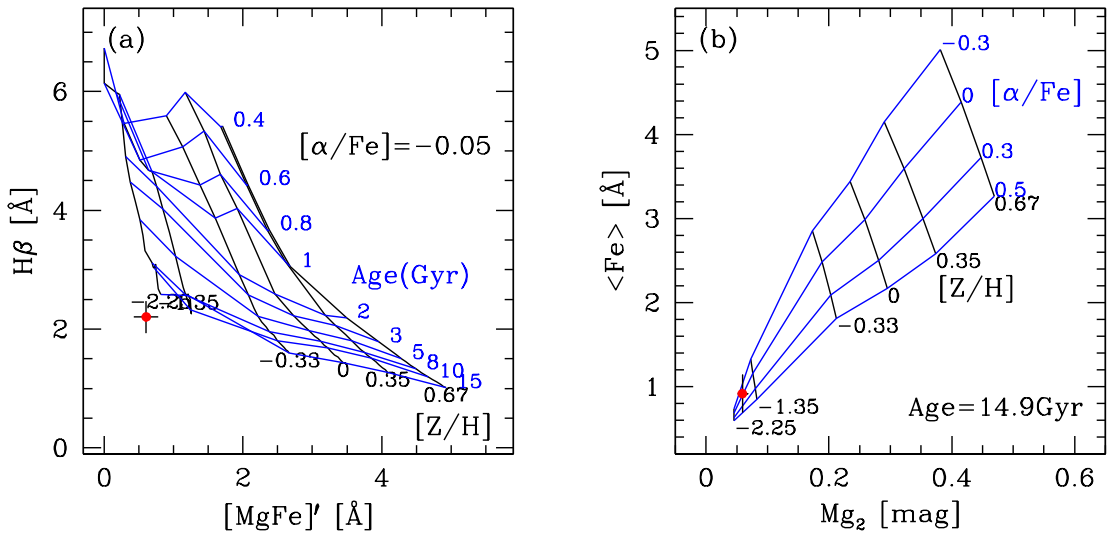}
\caption{(a) Lick line indices  for H$\beta$ versus [MgFe]$^\prime$ for GC-2, useful for measuring the metallicity and age (dots with errorbars).  
Grids represent the SSP models for various values of [Z/H] and age[Gyr] at [$\alpha$/Fe] = --0.05 given by \citet{tho03}.
(b) $<$Fe$>$ versus Mg$_2$ useful for measuring [$\alpha$/Fe]. The grids are for an age 14.9 Gyr.  
}
\label{figindexgrid}
\end{figure*}

  for Field-1, 
 $-3.95$  for GC-2, and 
$-3.98$  for Field-2.
Thus we finally calculate the distance modulus using $(m-M)_0 = I_{\rm 0, TRGB} - M_{\rm I, TRGB}$.
The distances derived are 
$(m-M)_0 = 27.80\pm 0.03$ ($d=3.63\pm0.05$ Mpc) for GC-1, 
$27.81\pm 0.03$ ($d=3.65\pm0.05$ Mpc) for Field-1,
$28.04\pm 0.04$ ($d=4.05\pm0.08$ Mpc) for GC-2, and 
$ 27.81\pm 0.03$ ($d=3.65\pm0.05$ Mpc) for Field-2.
This value is similar to the previous estimates for M81, $3.63\pm 0.14$ Mpc \citep{dur10,ger11}
and M82, $3.55\pm 0.11$ Mpc  \citep{lee12}. 
While GC-1 is at the same distance as M81, GC-2 is about 400 kpc behind the M81 halo along our line of sight.

\subsection{Spectral Line Analysis for GC-2} 

GC-2 was previously observed and classified as a galaxy in the SDSS
and its optical spectrum is available in the SDSS.
The spectrum of GC-2 shows several absorption lines typical for globular clusters.
We derived [Fe/H], [$\alpha$/Fe], and age for GC-2 from the comparison of Lick line index diagrams with the simple stellar population (SSP) models 
by \citet{tho03}.

Figure \ref{figindexgrid} displays H$\beta$ versus [MgFe]$^\prime$ diagram, and
 $<$Fe$>$ versus Mg$_2$ diagram for GC-2.
The composite index [MgFe]$^\prime$ is a combination of magnesium and iron-sensitive indices, 
defined as [MgFe]$^\prime = \sqrt{ {\rm Mgb(0.72\cdot Fe5270+0.28\cdot Fe5335)} }$. 
It is an excellent metallicity indicator, because it is completely independent of 
[$\alpha$/Fe], and this behavior is almost independent of the adopted age or metallicity \citep{tho03}. 
H$\beta$ is an efficient indicator for age. %
We used the line index data for these clusters provided by the SDSS. 
The values derived following the technique described in \citet{puz05} and \citet{par12}
are
[Fe/H] = --$2.3 \pm 0.12$,  [$\alpha$/Fe] = --$0.05 \pm 0.40$, and
 $\rm age=14.9 \pm 1.0$ Gyr.  This metallicity is consistent with the value derived from the color of the RGB. \rm
 These show that GC-2 is indeed very metal-poor and old.

\subsection{Spectral Energy Distribution Fit} 

We also derived age and mass from the spectral energy distribution fit using SDSS $ugriz$ magnitudes in comparison with the SSP model given by \citet{bru03}
($u=20.50\pm0.12$, $g=19.13\pm0.02$, $r=18.50\pm0.02$, $i=18.21\pm0.02$, and  $z=18.03\pm0.05$ for GC-1, and
 $u=19.51\pm0.06$, $g=18.25\pm0.01$, $r=17.60\pm0.01$, $i=17.31\pm0.01$, and  $z=17.10\pm0.03$ for GC-2).
Derived ages are log(age[y]) $ \sim10.2$ for both clusters as long as we adopt $Z = 0.0001$, and derived cluster masses are log($M/M_\odot )\sim6.40$ for GC-1, and 6.85 for GC-2. This age for GC-2 is also consistent with the value derived from the spectrum analysis.
Thus these clusters are very massive and old.

\begin{figure*}
\centering
\includegraphics[scale=0.85]{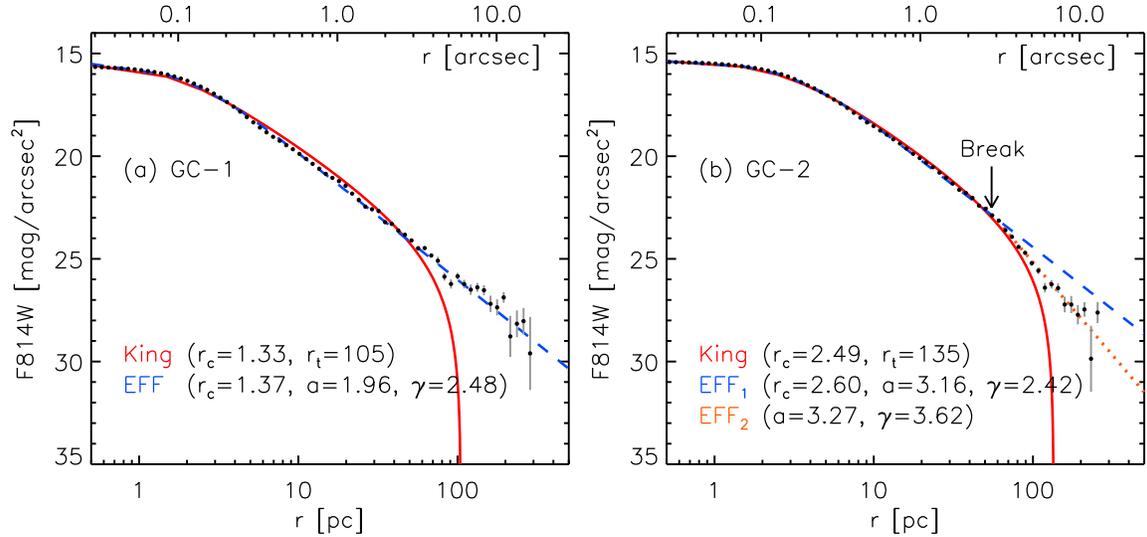}
\caption{$F814W$-band surface brightness profiles for GC-1 (a) and GC-2 (b) (dots with errorbars).
The thick solid lines represent the King model fit and the dotted and dashed lines represent the EFF power-law model fit.
}
\label{figsurf}
\end{figure*}

\subsection{Surface Photometry}

We derived surface photometry of GC-1 and GC-2 using IRAF/ELLIPSE task from the HST images. 
Figure 4  displays the radial profiles
of the surface brightness for $F814W$ images.
The radial profiles of the inner region look similar to the King profiles.
However, the outer parts in the radial profiles do not show any tidal cutoff, but continue to decrease smoothly.
From the King model fits for the inner region at 
$ 0\arcsec.0<r<4\arcsec.0$, we derived core radii  ($r_c$), 
$0\arcsec.0755 \pm 0\arcsec.0003$ ($1.329\pm0.005$ pc) for GC-1 and $0\arcsec.1269 \pm0\arcsec.0003$ ($2.491\pm0.006$ pc) for GC-2.
Also we fit the data using the model with a power law form in the outer region in \citet{els87},
$\sum (r_p ) = \sum_0 {( 1 + {r_p^2 \over a^2} )}^{-\gamma/2}$ 
where the scale radius $a$ is related with the King core radius  as
$r_c = a \sqrt {2^{2/\gamma} - 1 }$ for $r_t \gg r_c$.
There is a break at $r=2\arcsec.80$ (55 pc) for GC-2 so that we fit the data in two parts:
$a=1.96$, $\gamma=2.48$ for GC-1,
and  $a=3.16$, $\gamma=2.42$, $a=3.27$, $\gamma=3.62$ for GC-2.
The radial surface brightness profiles of GC-1 and GC-2 gradually decrease out to $r=15\arcsec$. 
By integrating the radial surface brightness profiles to $r=15\arcsec$, we derived integrated magnitudes of the star clusters, which we shall present in Section 4.1.
From these we derived half light radii ($r_h$) : $0\arcsec.352\pm0\arcsec.006$ ($6.13\pm0.01$ pc) for GC-1 and $0\arcsec.511\pm0\arcsec.005$ $(9.81\pm0.01$ pc) for GC-2.

\begin{figure}
\centering
\includegraphics[scale=0.95]{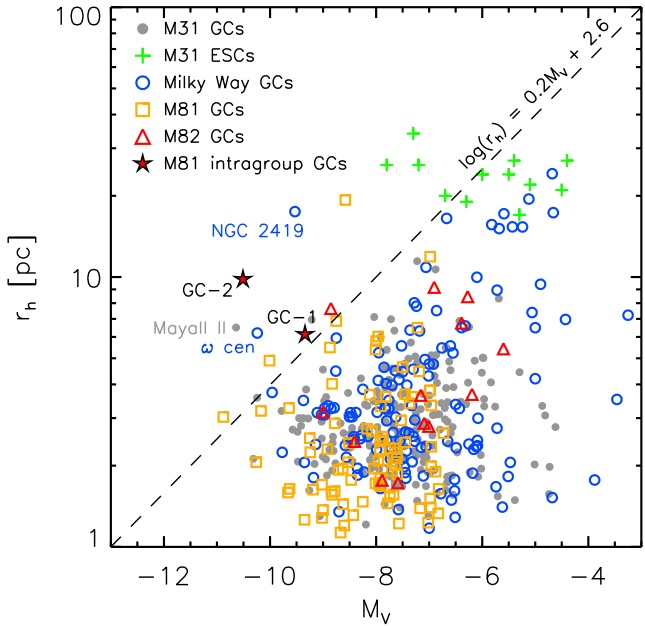}
\caption{Half light radii ($r_h$[pc]) versus $M_V$ for GC-1 and GC-2 (star symbols) in comparison with globular clusters in the Milky Way Galaxy (open circles) \citep{har96}, M31 (filled circles and crosses) \citep{hux09,hux11}, M81 (open squares) \citep{nan10,nan11}, and M82 (open triangles) \citep{lim12}.
}
\label{figrhmv}
\end{figure}

\section{Discussion and Summary}

\subsection{Luminosity-Size Relation}

Integrated magnitudes and colors of GC-1 and GC-2 are derived, respectively,
$V_0=18.46\pm0.01$ and $(V-I)_0=0.91\pm0.01$ and  $V_0=17.53\pm0.01$ and $(V-I)_0=0.90\pm0.01$.
Corresponding absolute magnitudes are
$M_{V} = -9.34\pm0.01$ for GC-1 and  $-10.51\pm0.01$ for GC-2, derived using our TRGB distances to the clusters.
In Figure \ref{figrhmv} we plot half light radii ($r_h$) versus absolute magnitude $M_{V}$ for GC-1 and GC-2 
in comparison with globular clusters in the Milky Way Galaxy \citep{har96}, M81
\citep{nan11} 
and M82 \citep{lim12}. 
We also plotted
the data for globular clusters and extended star clusters in M31 \citep{hux09,hux11}.

\citet{van04} presented a boundary relation between normal globular clusters and extended globular clusters: $\log (r_h) = 0.2 M_V + 2.6$.
GC-1 and GC-2 are found to be located above this boundary line,
like the case
of $\omega$ Cen and NGC 2419 in our Galaxy, and
Mayall II in M31. 
Thus GC-1 and GC-2 are larger than typical globular clusters with similar luminosity, and they are much brighter than the extended star clusters in M31.
The latter globular clusters ($\omega$ Cen, NGC 2419, and Mayall II) show some peculiar features so that they are often considered to be remnants of dwarf galaxies. 
Therefore GC-1 and GC-2 also may be in the same vein. 

\subsection{Intragroup Globular Clusters}

At the moment the most isolated globular cluster in the Local Group is known to be MGC1, located at the projected distance from M31, 117 kpc. Considering it is $\sim 160$ kpc closer than M31, its deprojected distance from M31 was derived to be  $200\pm20$ kpc \citep{mac10}.
It  is brighter ($M_V=-9.2$) than typical globular clusters, but fainter than GC-1 in M81.
In the Milky Way Galaxy, the most distant globular clusters and their  galactocentric distances are
NGC 2419 (91.5 kpc), Pal 3 (95.9 kpc),  Eridanus (95.2 kpc),  Pal 4 (111.8 kpc), and AM-1 (123.2 kpc) \citep{har96}.

On the other hand the most distant satellite dwarf galaxies in the two main Local Group galaxies are
Leo I in the Milky Way Galaxy and And XXVIII at $d>350$ kpc in M31 \citep{sla11}.
Leo I is a dwarf spheroidal galaxy located at 270 kpc, being considered long as the most distant satellite of the Milky Way Galaxy \citep{lee93b}. 
And XXVIII is a newly discovered dwarf galaxy about 100 kpc closer than M31. Its deprojected distance from M31 is estimated to be $365^{+17}_{-1}$ kpc
\citep{sla11}.
Thus the most remote globular clusters in our Galaxy and M31 are closer than the most distant satellite dwarf galaxies. 
This raises interesting questions:
{\it Can there be any globular clusters more distant than the most distant satellite galaxies in a galaxy? If so, what are they?}

GC-2 lies $406\pm97$ kpc behind M81 along our line of sight, according to our TRGB distance estimates to GC-2 and Field-2.
Considering this and the projected distance in the
sky we derive a three-dimensional distance from M81,  $408\pm97$ kpc.
Although it is closer to M82 than M81 in the sky, its three dimension distance is significantly larger than the projected distance from M82 (also from M81). Also M81 is twice as massive as M82 and is the most massive member in the M81 Group.
These indicate that  the main host for GC-2 may  be, if any, M81 rather
than M82.
Therefore GC-2 is the most isolated globular cluster among the known globular clusters in the local universe. 
In addition, it is  more distant than any known satellite dwarf galaxies around M81.
The radial velocity for GC-2 given in the SDSS  is $159\pm4$ km s$^{-1}$, which is much larger than the value for M81 ($-35\pm4$ km s$^{-1}$)  \citep{chy08}. 
 Thus GC-2 is receding with a velocity $\approx 200$ km s$^{-1}$ from M81.
 If this were moving at this velocity along the radial orbit, it would have taken about 2 Gyr to reach the current position from the center of M81.
Thus GC-2 can be considered as an intragroup globular cluster wandering
in the M81 Group. 

How can GC-2 be that far from M81, receding with a velocity of
$\approx 200$ km s$^{-1}$?   
The origin of GC-2 is not clear.
Possible scenarios are as follows. 
First, it might have ejected during the interaction of three main galaxies about 2 Gyr ago.
However, there are no evidence supporting this at the moment. 
Second, it may be one of the primordial globular clusters that formed early in isolation.
If so, where are others? 
Third, it may a remnant of a dwarf galaxy that accreted to M81 and is receding now.
If so, how could it survive during the perigalactic passage around M81?
These possibilities need to be investigated with observations or simulations.

\medskip

The authors thank anonymous referee for  useful suggestions. M.G.L. was supported in part by Mid-career Research Program through the NRF grant funded by the MEST (no.2010-0013875). I.S.J. and S.L. are grateful to Yeong Su Kim for financial support through the In Ha Kim scholarship. 
This paper is based on image data that obtained from Multimission Archive at the Space Telescope Science Institute (MAST).

\clearpage

\end{document}